\documentclass[nobm]{epl}
\usepackage[dvips]{epsfig,geometry}
\usepackage{amsmath}

\newcommand{\avr}[1]{\left\langle#1\right\rangle}
\newcommand{\figref}[1]{Fig. \ref{fig:#1}}


\title{Dynamics of short polymer chains
  in solution}

\author{A. Malevanets \and J. M. Yeomans}

\institute{Department of Physics, Theoretical Physics, 1 Keble Road,\\
  Oxford OX1 3NP, England}

\begin{document}

\maketitle

\begin{abstract}
We present numerical and analytical results describing the effect of
hydrodynamic interactions on the dynamics of a short polymer chain in
solution. A molecular dynamics algorithm for the polymer is coupled to
a direct simulation Monte Carlo algorithm for the solvent. We give an
explicit expression for the velocity autocorrelation function of the
centre of mass of the polymer which agrees well with numerical results if Brownian
dynamics, hydrodynamic correlations and sound wave scattering are
included.
\end{abstract}

\section{Introduction}

The aim of this paper is to present numerical and analytical results
describing the effect of hydrodynamic interactions on the dynamics of
a short polymer chain in solution.  Modelling a dilute polymer
solution is a difficult task.  This is because the dynamical behaviour
of the polymer can be dominated by hydrodynamic interactions between
different parts of the polymer chain.  These interactions are
long-ranged and develop very slowly compared to the time scale of the
Brownian fluctuations of individual monomers. Thus if a molecular
dynamics time-step is chosen to correctly follow the Brownian dynamics
it becomes prohibitively expensive in computer time to also model the
dynamics of enough solvent molecules for sufficient time to allow
hydrodynamic correlations to develop.

To move towards overcoming this problem we introduce a hybrid
numerical approach. The dynamics of the chain is treated exactly
(within the limitations of a finite time step) by a molecular dynamics
solution of Newton's equations of motion. The solvent is modelled
using a direct simulation Monte Carlo algorithm\cite{bird:mole}.  The
algorithm is constructed in such a way that energy and momentum are
conserved and that at equilibrium the system is described by a
microcanonical distribution.  However, molecular details of the
solvent are not included: it acts as a momentum-conserving heat bath
which can support hydrodynamic modes.

The physical quantity we choose to measure in the simulations is the
velocity autocorrelation function of the centre of mass motion of a
polymer chain.  This provides clear evidence of a fast exponential
decay at early times which results from collisions in which the
solvent particles are uncorrelated and a slow algebraic decay at later
times arising from hydrodynamic interactions that develop in the
fluid.  We propose a theory which reproduces the behaviour of the
velocity autocorrelation function and demonstrate that both limits
arise naturally from a simple equation of motion which incorporates
the solvent--polymer interaction as a viscous drag term. We also argue
that, as the fluid is compressible, a feature common to many mesoscale
approaches, both transverse and longitudinal velocity modes
participate in momentum transfer yielding a significant correction to
the value of the chain diffusion coefficient\cite{zwanzig70:hydrod}. 
The theoretical prediction
is shown to agree well with the numerical results.

The paper starts with a description of the numerical method. Next we
present an analytic approach to the calculation of the velocity
autocorrelation function. We emphasise that both the Brownian and the
hydrodynamic contributions are important, and that the latter includes
contributions from sound waves.  We show how the Kirkwood formula for
the diffusion constant of the centre of mass of the polymer chain
follows from the theoretical development and hence discuss the
relation of our work to previous mesoscale simulations of polymer
dynamics\cite{kong97:effec,ahlrichs99:simul,spenley}.

\section{Direct simulation Monte-Carlo  algorithm}

\label{dsmc}

We construct a hybrid algorithm for the dynamics of polymers in
solution by exploiting the time-scale separation between the
microscopic motion of the polymer beads and the much slower
propagation of hydrodynamic modes within the fluid. The evolution of
the polymer beads follows a molecular dynamics algorithm.  The solvent
is modelled on a mesoscopic length scale using a direct simulation
Monte Carlo approach.

Consider first the solvent. The direct simulation Monte Carlo
algorithm is based upon the motion of $N_s$ particles in continuous
space but discrete time. Let the particles, labelled $i=1,2\ldots
N_s$, take positions $x_i(t)$ with velocity $u_i(t)$ at time $t$.  The
evolution for unit time comprises a free streaming step applied to
each particle $i$
\begin{equation}
  \label{eq:free-str}
  \mathbf{x}_i(t+1) =  \mathbf{x}_i(t) + \mathbf{u}_i(t)
\end{equation}
followed by a collision step.  The collision step must (i) preserve
mass, momentum and energy locally so that the macroscopic equations of
motion are obtained in the continuum limit; (ii) allow momentum
transfer between the particles so that equilibrium can be achieved.
Many different schemes are possible. Here we follow Malevanets and
Kapral\cite{malevanets99:mesos} and divide space into $L_x\times
L_y\times L_z$ cubic cells each containing, on average, $N_s/L_x L_y
L_z$ particles.  The secular velocities of particles in each cell are
then rotated according to
\begin{equation}
  \label{eq:dsmc-rot}
  \mathbf{u}_i(t+1) = \mathbf{U}(t) + 
  \mathcal{R}(\mathbf{u}_i(t) - \mathbf{U}(t))
  \mbox{ ,}
\end{equation}
where $ \mathbf{U}$ is the velocity of centre of mass of particles in
each cell and $\mathcal{R}$ is a rotation matrix. In the present study
$\mathcal{R}$ is taken to be a rotation by $\pi/2$ around an
arbitrarily chosen axis.

Various hydrodynamic properties of the solvent are needed to compare
the analytical and numerical results. These depend on the details of
the collision step and are most easily obtained numerically using
either the Green-Kubo formalism or non-equilibrium dynamics
simulations.  Parameters for the solvent component are temperature
$k_BT = 1.0$, mass $m=4.0$ and density $\rho=2.0$.  A time integral
over the stress autocorrelation function gives the kinematic viscosity
coefficient as $\nu = 0.25$.  The decay rate of a small longitudinal
sinusoidal perturbation gives the attenuation rate of a sound wave to
be $ \Gamma = 0.275$ and the speed of sound $c = 0.645$.

A chain is introduced into the system by choosing $N$ beads which are
connected by the harmonic potential $W_n = \kappa \|\mathbf{r}_n -
\mathbf{r}_{n+1} \|^2/2$. Apart from the harmonic potential the
polymer beads are taken to be non-interacting to facilitate comparison
to the predictions of the Zimm and Rouse theories\cite{doi96}. The
beads are included in the same collision step \eqref{eq:dsmc-rot} as
the solvent particles. In between the collision steps the polymer
dynamics is integrated using a velocity Verlet
algorithm\cite{tuckerman92:rever} with time step $\Delta t$.

\begin{figure}[htbp]
  \begin{center}
    \epsfig{file=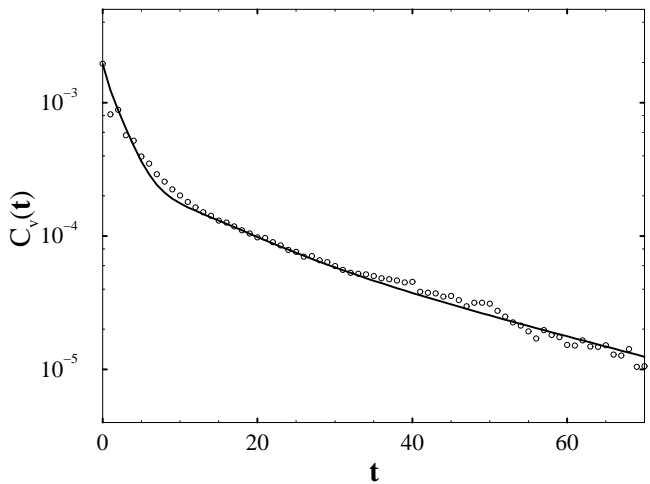,width=8.5cm}
    \caption{
      Velocity autocorrelation function of a polymer chain.  The
      circles show the computed values and the solid
      line represents the theoretical curve calculated using equation
      \protect\eqref{eq:vacf}. }
    \label{fig:vacf}
  \end{center}
\end{figure}

The simulations were used to measure the velocity
autocorrelation function of the centre of mass of the polymer chain at
equilibrium. The results are presented in \figref{vacf}. The
parameters used in simulations were $M=8.0$, $N=128$ and $\kappa =
3.0$. The radius of gyration of the chain was $R_g =
\sqrt{\frac{Nb^2}{6}} = 2.31$.

\section{Calculation of the centre of mass velocity autocorrelation function}

The aim of this section is to derive a formula for the velocity
autocorrelation function of the centre of mass of the polymer chain.
We commence by writing the equation of motion
of the monomers as\cite{hansen86:theor} 
\begin{equation}
  \label{eq:mon-dyn}
  M\frac{d}{dt}\mathbf{v}_n(t) = \mathbf{f}_n(t) 
  + \xi(\mathbf{u}(\mathbf{r}_n(t),t) - \mathbf{v}_n(t)) \mbox{ ,}
\end{equation}
where $\mathbf{v}_n$ is the velocity of $n$th monomer and
$\mathbf{r}_n$ its position at time $t$. $\mathbf{f}_n$ is the force
on the $n$th monomer exerted by the other beads in the chain.
$\mathbf{u}(\mathbf{r}_n(t),t)$ is the velocity of the solvent at
$\mathbf{r}_n$ at time $t$. The coefficient $\xi$ gauges the strength
of viscous friction between the chain and the solvent.

We are interested in the behaviour of the velocity of the centre of
mass of the polymer chain $\mathbf{V}(t)$. Summing \eqref{eq:mon-dyn}
over all beads, we note that the contribution from the internal forces
cancel. The resulting equation can be integrated to give
\begin{equation}
  \label{eq:cm-evol}
  \mathbf{V}(t) = e^{-\xi t/M} \mathbf{V}(0) + 
  \frac{\xi}{MN}\int\limits_0^t 
  d\tau e^{-\xi (t-\tau)/M} \sum_{n} \mathbf{u}(\mathbf{r}_n(\tau),\tau)
  \text{ .}
\end{equation}

The next task is to calculate $\mathbf{u}(\mathbf{r}_n(\tau),\tau)$.
This can be done by using the linearized thermo-hydrodynamic equations
for a compressible fluid\cite{boon:mole}. We assume the transfer of
momentum from the chain to the fluid is sufficiently fast that
\begin{equation}
  \label{eq:init-cond}
  \mathbf{u}(\mathbf{x},0) = \frac{M}{\rho}\sum_{n} \mathbf{v}_n 
  \delta(\mathbf{x}-\mathbf{r}_n(0)) \mbox{ .}
\end{equation}
Evolution of the fluid velocity field is then given by\cite{boon:mole}
\begin{equation}
  \label{eq:v-evol}
  u_{\alpha}(\mathbf{k},t) = \sum\limits_n \frac{M}{\rho}
  \left[ \frac{1}{k^2}(k^2 \delta_{\alpha\beta} - 
    k_{\alpha}k_{\beta})e^{-\nu k^2 t} + 
  \frac{1}{k^2}k_{\alpha}k_{\beta} e^{-\Gamma k^2 t}
  \cos{kct} \right]e^{i\mathbf{k}\cdot\mathbf{r}_n(0)} 
v_{n\beta}(0) \mbox{ .}
\end{equation}

Using the identity
\begin{equation}
  \label{eq:ur}
  \mathbf{u}(\mathbf{r}_n(t),t) = \frac{1}{(2\pi)^3} 
  \int d\mathbf{k} e^{-i\mathbf{k} \cdot \mathbf{r}_n(t)} \mathbf{u}(\mathbf{k},t)
  \text{ ,}
\end{equation}
we can now calculate the chain centre of mass velocity autocorrelation
function
\begin{equation}
  \label{eq:def-vacf}
  C(t) = \frac{1}{3} \avr{\mathbf{V}(t)\mathbf{V}(0)} \mbox{ ,}
\end{equation}
where the angular brackets denote an equilibrium ensemble average over
the Gibbs distribution.  Substituting the expression for the fluid
velocity \eqref{eq:v-evol} into the formula for the polymer centre
of mass velocity \eqref{eq:cm-evol} and using the result in \eqref{eq:def-vacf} gives
\begin{equation}
  \label{eq:vacf}
  C(t) = e^{-\xi t/M} \frac{k_B T}{MN} + 
  \frac{k_BT \xi}{\rho M (2\pi)^3 N } 
  \int\limits_0^t d\tau e^{-\xi (t-\tau)/M} 
  \int d\mathbf{k} \left[ \frac{2}{3} e^{-\nu k^2 \tau} + 
  \frac{1}{3}e^{-\Gamma k^2 \tau}
  \cos{kc \tau} \right] S(\mathbf{k},\tau) \mbox{ ,}
\end{equation}
where we have used the usual definition of the structure factor
\begin{equation}
  \label{eq:def-sf}
   S(\mathbf{k},\tau) =  \frac{1}{N} \avr{\sum\limits_{l,n}
     e^{i\mathbf{k}\cdot(\mathbf{r}_n(\tau)-\mathbf{r}_l(0))}} \text{ .}
\end{equation}

We assume that the structure factor exhibits a two-stage decay;
an initial fast decay due to the diffusion of individual beads 
followed by a later stage when it decays as
$\exp(-Dk^2t)$.  We found that in our simulations the structure factor
is well approximated by
\begin{equation}
  \label{eq:sf}
  S(\mathbf{k},t) = N e^{-Dk^2t} e^{-R_gk^2/3} \mbox{ ,}
\end{equation}
where $D \ll \nu$. This expression is obtained assuming that the
chain configurations at times $0$ and $t$ are uncorrelated.

The behaviour of $C(t) $ obtained from a numerical integration of
equation \eqref{eq:vacf} using \eqref{eq:sf} is compared to
the simulation results in \figref{vacf}. Agreement is good. Small
deviations of the theoretical curve from the numerical results on
short-times scales can most probably be attributed to a breakdown of
the assumption of a single friction coefficient with no memory effects
in equation \eqref{eq:mon-dyn}.

It is sometimes helpful to divide $C(t)$ into a
short time or Brownian and a long time or hydrodynamic contribution.
These can be identified as the first and second terms in equation
\eqref{eq:vacf}. Note that there is no clear time scale separation
between the two contributions: we expect this to be the case for the
short polymers considered in mesoscale simulations. Note also that
both contributions follow from the same physical mechanism, the
viscous drag between the fluid and the polymer. However, the first
term corresponds to Brownian motion when the fluid molecules can be
considered to move at random; the second occurs when hydrodynamic
correlations in the fluid lead to interactions between the monomers.
It can be ``switched off'' in the simulations by randomizing the
solvent molecules' velocities after each collision step.

The second term in the square brackets in \eqref{eq:vacf} arises from
sound wave--chain interactions. This coupling may lead to a negative
minimum in the velocity autocorrelation
function\cite{zwanzig70:hydrod}.

\section{Calculations of the chain centre of mass diffusion coefficient}

In this section we estimate contributions from different terms of
equation \eqref{eq:vacf} to the diffusion coefficient of the centre of
mass of the polymer chain.
Replacing $S(\mathbf{k},t)$ in equation \eqref{eq:vacf} by its value
at some intermediate time $S(\mathbf{k},t')$ and
integrating gives
\begin{equation}
  \label{eq:diff-notime}
  D = \frac{k_BT}{N\xi} + \frac{1}{N^2}
  \sum\limits_{l,n}\left\{\frac{k_B T}{6\pi\eta}
  \avr{\frac{1}{|\mathbf{r}_l(t)-\mathbf{r}_n(t')|}} +
  \frac{k_B T}{12\pi\rho\Gamma}
  \avr{\frac{e^{-c|\mathbf{r}_l(t)-\mathbf{r}_n(t')|/\Gamma}}
    {|\mathbf{r}_l(t)-\mathbf{r}_n(t')|}} \right\}
  \text{ .}
\end{equation}
In equation \eqref{eq:diff-notime} the first term corresponds to the
microscopic or uncorrelated contribution and has the same scaling as
Rouse's model predicts. The second term gives the Zimm contribution
due to hydrodynamic interactions between beads and the last term
arises from scattering of sound waves on the chain as the
longitudinal component of momentum drains from the system.

For the simulations the first and third terms in this equation are
approximately $10\%$ of the total value. The same order corrections
are expected in water, where typical parameters are $c \sim 10^3$ ms$^{-1}$
and $\Gamma \sim 10^{-6}$ m$^2$s$^{-1}$, for chains with radius of
gyration $R_g \sim 10^{-8}$m. For longer chains when the condition $R_g c
\gg \Gamma$ is satisfied the sound contribution to the diffusion
coefficient vanishes.

\begin{figure}[htbp]
  \begin{center}
    \epsfig{file=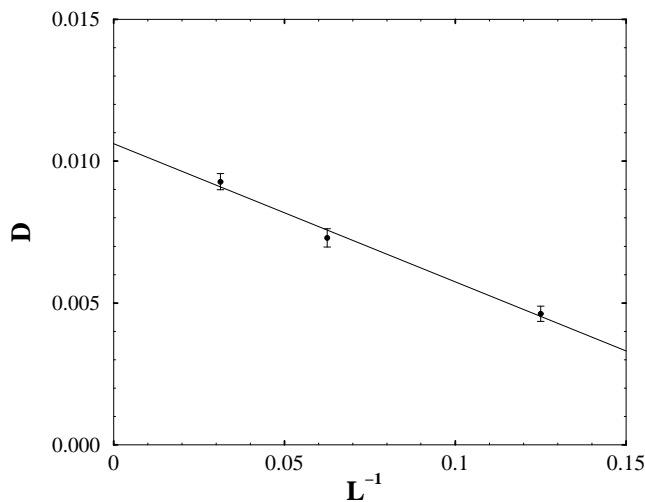,width=8.5cm}
    \caption{
      Diffusion coefficient as a function of system size. Solid
      circles correspond to numerical values of the diffusion
      coefficient for systems of size $8$, $16$ and $32$. The line
      is a fit to the data.}
    \label{fig:dofl}
  \end{center}
\end{figure}

Hydrodynamic interactions are long range and finite-size effects lead
to large corrections to the diffusion
coefficient\cite{dunweg93:molec}.  In order to estimate this effect we
replace integration over $\mathbf{k}$ in equation \eqref{eq:vacf} with
the summation over discrete wave-vectors
$\mathbf{k}=(2\pi/L)\mathbf{n},k\neq 0 $. Then in the
intermediate time regime $ M/\xi \ll t \ll R_g^2/\nu$ we obtain
\begin{equation}
  \label{eq:L-vacf}
  C_L(t) = C(t) - \frac{k_BT}{\rho (2\pi)^3 N} \int\limits_{|k|_1<\pi/L} 
  d\mathbf{k}\frac{2}{3} S(\mathbf{k},t)
\end{equation}
Hence, assuming that the structure factor varies only weakly with time,
\begin{equation}
  \label{eq:d-finite}
  D_L = D - \frac{2}{3}\frac{k_BT}{(2\pi)^3\eta} \int\limits_{|k|_1<\pi/L}
  \frac{1}{k^2}d\mathbf{k} = \frac{k_BT}{6\pi\eta}\left[ \frac{1}{R_h} -
  \frac{2.44}{L}\right] \mbox{ .}
\end{equation}
This form of system size dependence is consistent with the correction
obtained by Dunweg \emph{et al.}\cite{dunweg93:molec} using different
arguments.  In \figref{dofl} we present values of the diffusion
coefficient of a chain obtained in simulations of systems with
dimensions $8^3$, $16^3$ and $32^3$. A numerical fit gives for the
finite size correction
\begin{equation}
  \label{eq:d-exp}
  D_L = D - \frac{k_BT}{6\pi\eta} \frac{1.84}{L}
\end{equation}
where $D = 1.05\times 10^{-2}$. Alternately we may compute the
hydrodynamic contribution to the
diffusion coefficient from equation \eqref{eq:diff-notime}. Averaging
yields
\begin{equation}
  \label{eq:diff-approx}
  D = \frac{k_BT}{6\pi\eta \sqrt{\pi/3} R_g} = 1.15\times 10^{-2} \mbox{ .}
\end{equation}
This is consistent with the numerical value of the diffusion coefficient obtained in the
simulations within
statistical errors and the difficulties inherent in calculating the finite-size corrections.

\section{Discussion}

We have described a new mesoscale method to simulate the dynamics of
polymer chains in solution. The polymer chain itself is modelled using
standard molecular dynamics. This has the advantage that the
architecture of the chain and the interactions within it can be
varied. The solvent is described by a direct simulation Monte Carlo
algorithm which reproduces the thermohydrodynamic equations of motion
on sufficiently long length scales.  By ignoring the molecular detail
of the solvent it becomes feasible to study the build up of
hydrodynamic correlations between different monomers.

The numerical results were shown to agree well with an equation of
motion for the monomers which assumed a linear coupling to the
velocity of the surrounding fluid via a friction coefficient $\xi$.
The velocity-velocity correlation function comprises two
contributions; an initial exponential decay resulting from Brownian
collisions of monomers with uncorrelated solvent molecules and, at
longer times, a hydrodynamic contribution as monomers interact via
hydrodynamic modes propogated through the solvent.  There is no clear
time-scale separation between the two contributions for polymer chain
lengths accessible to computer simulations. Hence the Zimm limit,
which assumes that the velocity-velocity correlation function has
decayed to zero is not appropriate for the systems we consider here.
However a Kirkwood formula for the diffusion constant
\eqref{eq:diff-notime} follows naturally and shows that the Zimm
scaling $D \sim R_G^{-1}$ for the hydrodynamic contribution to the
diffusion constant is still expected.

Other mesoscale approaches which have been used to study dilute
polymer solutions include dissipative particle
dynamics\cite{kong97:effec,spenley} and a polymer chain updated using
molecular dynamics coupled to a lattice Boltzmann
solvent\cite{ahlrichs99:simul}. Both approaches gave good results for
the Zimm scaling of the diffusion constant.  Qualitative values were
consistent but precision was limited by the difficulty in separating
out the Brownian contribution from the hydrodynamic one and because of
strong finite-size corrections. It would be of interest to measure
velocity--velocity correlation functions using these methods to enable
a more detailed comparision of the dynamics to that obtained using the
mesoscale approach described here.

Several open questions remain. We have concentrated on the dynamics of
the centre of mass of the polymer chain. It would be interesting to
next consider the internal dynamics (Rouse modes) which will be
affected by the interaction between the polymer and the solvent on
shorter length and time scales. It will also be possible to
investigate polymer dynamics under shear and the effect of the
architecture of the polymer chains on their flow behaviour.

\section{Acknowledgements} We have benefitted greatly 
from discussions with B.  Dunweg, C. Manke, N.A. Spenley and P. Ahlrichs.

%\bibliographystyle{unsrt}
%\bibliography{/home/wytham/amalevan/work/appl/uwo/cv,/home/wytham/amalevan/work/paper/jcp,/home/wytham/amalevan/doc/thesis/thesis}

\begin{thebibliography}{10}
  
\bibitem{bird:mole} G.~A. Bird.  \newblock {\em Molecular gas
    dynamics}.  \newblock Oxford University Press, London, 1976.
  
\bibitem{zwanzig70:hydrod} R.~Zwanzig and M.~Bixon.  \newblock
  Hydrodynamic theory of the velocity correlation function.  \newblock
  {\em Phys. Rev. A}, 2(5):2005, 1970.
    
\bibitem{kong97:effec} Y.~Kong, C.~W. Manke, W.~G. Madden, and A.~G.
  Schlijper.  \newblock Effect of solvent quality on the conformation
  and relaxation of polymers via dissipative particle dynamics.
  \newblock {\em J. Chem. Phys.}, 107(2):592, 1997.
  
\bibitem{ahlrichs99:simul} P. Ahlrichs and B. Dunweg.  \newblock
  Simulation of a single polymer chain in solution by combining
  lattice Boltzmann and molecular dynamics.  \newblock {\em J. Chem.
    Phys.}, 111(17):8225, 1999.
\bibitem{spenley} N.~A. Spenley. \newblock Scaling laws for polymers in
dissipative particle dynamics. \newblock {\em Europhys. Lett.}, in press.
\bibitem{malevanets99:mesos} A. Malevanets and R. Kapral.  \newblock
  Mesoscopic model for solvent dynamics.  \newblock {\em J. Chem.
    Phys.}, 110(17):8605, 1999.
  
\bibitem{doi96} M.~Doi.  \newblock {\em Introduction to polymer
    physics}.  \newblock Clarendon Press, Oxford, 1996.
  
\bibitem{tuckerman92:rever} M.~Tuckerman, B.~J. Berne, and G.~J.
  Martyna.  \newblock Reversible multiple time scale molecular
  dynamics.  \newblock {\em J. Chem. Phys.}, 97(3):1990, 1992.
  
\bibitem{hansen86:theor} J.-P. Hansen and I.~R. McDonald.  \newblock
  {\em Theory of simple liquids}, chapter 8.7.  \newblock Academic
  Press, 1986.
  
\bibitem{boon:mole} J.~P. Boon and S. Yip.  \newblock {\em Molecular
    hydrodynamics}.  \newblock McGraw-Hill, New York, 1980.
  
\bibitem{dunweg93:molec} B.~Dunweg and K.~Kremer.  \newblock
  Molecular-dynamics simulation of a polymer-chain in solution.
  \newblock {\em J. Chem. Phys.}, 99(9):6983, 1993.
  

\end{thebibliography}

\end{document}